# Malaria detection in Segmented Blood Cell using Convolutional Neural Networks and Canny Edge Detection

Tahsinur Rahman Talukdar[1,*], Mohammad Jaber Hossain[1], Tahmid H. Talukdar[2]

[1]*Department of Computer Science and Engineering, Leading University, Sylhet, Bangladesh.*
[2]*Holcombe Department of Electrical and Computer Engineering, Clemson University, USA*
*\*corresponding author. email: tahsinurrahmantalukdar@gmail.com*

Abstract: *We apply convolutional neural networks to identify between malaria infected and non-infected segmented cells from the thin blood smear slide images. We optimize our model to find over 95% accuracy in malaria cell detection. We also apply Canny image processing to reduce training file size while maintaining comparable accuracy (~ 94%).*

## Introduction

Conventional optimal methods of infection diagnosis required sluggish, laborious, costly, and professional expertise. Automated diagnosis solutions provide an economical, swift, efficient, and accurate alternative. Computer vision based change detection is useful for various number of applications in multiple sectors [1]. Medical imaging is an influential advance in the field of medicine [2]. Especially, digital image processing for medical applications is an emerging topic of interest [3]. An automated process for microbiological identification is an optimistic solution for reducing analysis time and reduce workload on medical personnel [4]. Machine learning and neural networks can have a strong impact in alleviating these workloads. It has already shown success in some medical fields [5]. Neural networks are becoming more popular in use as they are more frequently being applied for pattern recognition and classification in images [6].

Image classification is important for computer vision which has significance in our everyday life [7]. In particular, facial recognition is currently ubiquitous thanks to widespread use of social media [8]. Inspiration for convolutional neural networks (CNN) comes from the mammalian visual system structure [9]. CNNs are gaining more wide-spread use in phones which can lead to large commercial application in the medical field [10]. It is domineering the field of machine learning approach for visual object recognition [11]. CNN as a machine learning algorithm is a fast classifier of big data as well as accurate predictor of disease[12]. It has revolutionized pattern recognition of images which includes medical images [13].

In general, CNNs have deeper layers and are arranged in a volumetric fashion with height, width as well as depth [14]. It has shown great results in computer vision tasks[15]. CNNs is a popular machine learning algorithm not only in the computer vision community but also natural language processing to hyperspectral image processing [16], [17] and to medical image analysis [18].

# Methodology

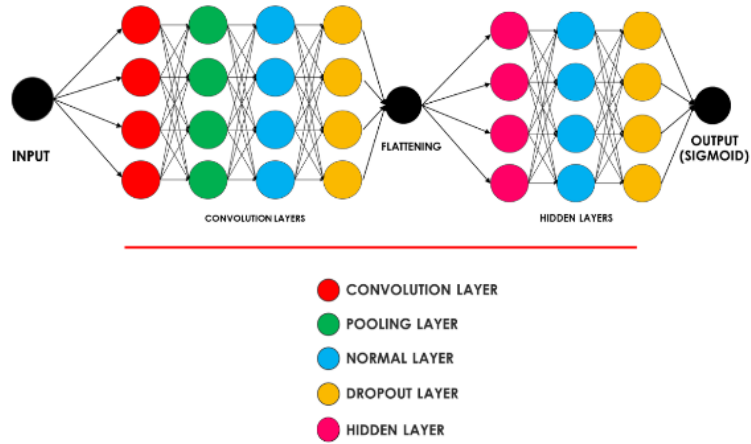

Fig 1. Overview of the convolutional neural network (CNN) model implemented in this work

Our dataset is provided by National Institute of Health (NIH). It is a repository containing thin blood smear slide images from the *Malaria Screener* research activity [19]. Giemsa-stained thin blood smear slides were collected from 150 infected and 50 healthy patients in Chittagong Medical College Hospital, Bangladesh. The images are captured using a smartphone camera in the microscopic view. Images were classified manually by an expert slide reader at Mahidol-Oxford Tropical Medicine Research Unit in Bangkok, Thailand. The dataset contains a total of 27,558 cell images with equal amounts of infected and un-infected cells [19].

We separate the images in two folders and train our convolutional neural networks to classify them. As illustrated in Fig. 1, our CNN has 4 parts: (1) Convolution layer, (2) Pooling Layer, (3) Normal Layer, and (4) Dropout layer.

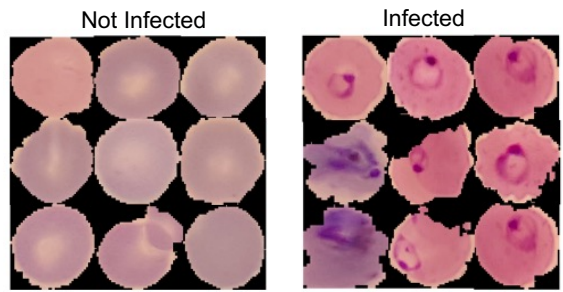

Figure 2. Not infected vs infected cell images

Example input images are shown in Figure 2. There is a clear difference when observed manually which we expect the neural networks to pick up fairly accurately. Our convolution layer inputs the images based on the sizes sent to the image (64x64 in this case). A 3x3 kernel filter is applied to the original 64x64 image. The Pooling layer to be a more specific max-pooling layer picks the maximum number in each patch in the feature map in the matrix. In this case, the max pool feature map is (2x2).

Batch normalization makes each mini-pact ready for the inputs to a layer for which steadies the learning process and minimizes the number of training epochs needed to train neural networks. Dropout means dropping neurons randomly in the neural network to prevent overfitting. At first the efficiency was 80%. After adding more layers, the efficiency increases to 85-90%. After further epoch increase to 95%. Drawback being that increased as well. An extra layer being flattening layer that creates a long line of input data in vector form for the neural network.

Our CNN includes a hidden layer which consists of 3 parts: (1) Hidden Layer, (2) Normal layer, and (3) Dropout layer. The dense layer contains interconnected 512 neurons. The entire three parts make up 1 entire Hidden Layer. We implemented 4 CNN and 4 Hidden Layer networks.

Now at the end of processing activation function determines the result. We implemented a sigmoid function here. However, "tanh" and "relu" show similar classification performances. Adam optimizer is used because its common and training cost effective. Categorical cross-entropy is common. We find maximum accuracy is reached around epoch is 4-5. Further increase in number of epochs results in higher training time but without any increase in accuracy. Lower number of epochs are also preferable to avoid overfitting. We find that the best approach is to keep an epoch which minimizes training time and maximizes tested accuracy. This process will be repeated until the accuracy of the process reaches its optimal state.

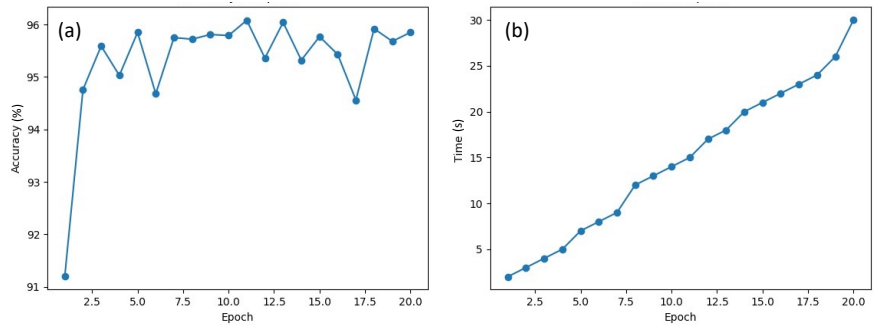

Figure 3. (a) Accuracy vs epoch (b) training time vs epoch

We further processed images via Canny edge detection and removed RGB colors only left edges which reduced size for better memory utilization in database centers. Removing colors slightly reduce the accuracy to 94%. Training time remains constant regardless. Altered images have $1/3^{rd}$ the size of the original. Canny edge detection has lower granules so the possibilities for false positive results are lower that's why this edge detection method was prioritized. Example image where Canny edge detection is applied is shown in Figure 4.

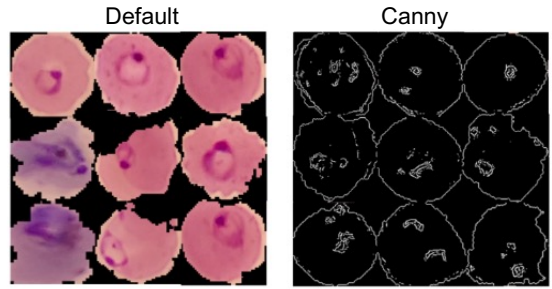

Figure 4. Default vs Canny edge detection algorithm and output image

Image edge contains internal information like direction, characteristics, shape etc, and its also widely used in image segmentation, categorization, registration, and pattern recognition [20]. Edge detection is an efficient and effective processing tool of images which provides necessary information and characteristics for those images [21]. Medical images edge detection is important for object recognition of the human organs [22]. Data classification in presence of noise leads to less accurate results [23]. Optimized and skillfully designed CNNs have potential to provide a better performance in the future [24].

## Results:

Results are illustrated in Table 1.

Table 1. Results

|  | Raw | CANNY |
|---|---|---|
| *Accuracy (Test1, Test2, Test3)* | 95.43% | 94.52% |
|  | 95.88% | 94.78% |
|  | 95.45% | 94.36% |
| *Training Time (in seconds)* | 584s | 586s |
|  | 572s | 578s |
|  | 573s | 594s |
| *Image Size(in Megabytes)* | 334 MB | 139 MB |
| *Epoch number* | 5 ||
| *Image Numbers* | 27558 Images ||
| *System* | Intel(R) Core i7, 12 GB ||

## Accuracy

Average accuracy: 95%
Canny Edge Detection accuracy: 94%

## Data size

Default Data Size: 334 MB.
Canny Edge Detection Data Size: 139 MB. (1/3$^{rd}$ of Default Data Size)

## Training time

Default Training Time: 9-10 Minutes
Canny Edge Detection Training Time: 9-10 Minutes

Improvements in accuracy can be made by using more images in datasets.

## Conclusion

In conclusion, we demonstrate a convolutional neural network classifier that is able to identify malaria infected cells from microscope slide images with near 95% accuracy with a training time of less than 10 minutes. We believe this would enable a smartphone based automated process of fast and efficient diagnosis while reducing the workload of healthcare workers. This can impact the way we diagnose and detect and store data for viruses in the future.